\PassOptionsToPackage{dvipdfmx}{graphicx}
\documentclass[lettersize,journal]{IEEEtran}
\usepackage{amsmath,amsfonts}
\usepackage{algorithmic}
\usepackage{array}
\usepackage[caption=false,font=normalsize,labelfont=sf,textfont=sf]{subfig}
\usepackage{textcomp}
\usepackage{stfloats}
\usepackage{url}
\usepackage{verbatim}
\usepackage{graphicx}
\usepackage{cite}
\usepackage{threeparttable}
\usepackage{booktabs}
\usepackage{multirow}
\usepackage{algorithm}
\usepackage{caption}
\usepackage{siunitx}
\usepackage{amssymb}
\usepackage{bm}

\newcommand{\jj}{\mathrm{j}}
\newcommand{\dd}{\mathrm{d}}
\def\BibTeX{{\rm B\kern-.05em{\sc i\kern-.025em b}\kern-.08em
    T\kern-.1667em\lower.7ex\hbox{E}\kern-.125emX}}
\usepackage{balance}
\begin{document}
\title{Simulation for Noncontact Radar-Based Physiological Sensing Using Depth-Camera-Derived Human 3D Model with Electromagnetic Scattering Analysis}
\author{Kimitaka~Sumi,~\IEEEmembership{Graduate Student Member,~IEEE},
  Takuya~Sakamoto,~\IEEEmembership{Senior Member,~IEEE}
  \thanks{
    The authors are with the Department of Electrical Engineering, Graduate School of Engineering, Kyoto University, Kyoto 615-8510, Japan (e-mail: sumi.kimitaka.25r@st.kyoto-u.ac.jp; sakamoto.takuya.8n@kyoto-u.ac.jp).
  }}

% \markboth{Simulation for Noncontact Radar-Based Physiological Sensing Using Depth-Camera-Derived Human 3D Model}
% {Simulation for Noncontact Radar-Based Physiological Sensing Using Depth-Camera-Derived Human 3D Model}

\maketitle

\begin{abstract}
  This study proposes a method for simulating signals received by frequency-modulated continuous-wave radar during respiratory monitoring, using human body geometry and displacement data acquired via a depth camera. Unlike previous studies that rely on simplified models of body geometry or displacement, the proposed approach models high-frequency scattering centers based on realistic depth-camera-measured body shapes and motions. Experiments were conducted with six participants under varying conditions, including varying target distances, seating orientations, and radar types, with simultaneous acquisition from the radar and depth camera. Relative to conventional model-based methods, the proposed technique achieved improvements of 7.5\%, 58.2\%, and 3.2\% in the correlation coefficients of radar images, displacements, and spectrograms, respectively. This work contributes to the generation of radar-based physiological datasets through simulation and enhances our understanding of factors affecting the accuracy of non-contact sensing.
\end{abstract}


\section{Introduction}
\IEEEPARstart{P}{hysiological} signals such as respiration and heartbeat signals are critical indicators of respiratory and cardiovascular conditions, making their monitoring essential \cite{10.1001/jama.1980.03310100041029, 10.1007/BF02600071, 10.1016/S1520-765X(03)90001-0, 10.1016/j.resuscitation.2006.08.020, 0002-838X}. Continuous monitoring of these signals in infants, seniors, and patients with sleep apnea can allow early detection of diseases and reduce the burden on medical and care systems \cite{10.2196/18636}. Although physiological signals are typically measured using contact sensors, non-contact techniques are gaining traction because of concerns regarding user discomfort and the desire for improved ease of use. Common noncontact methods include the use of RGB cameras \cite{10.1364/OE.18.010762, 10.1364/BOE.8.004838}, acoustic sensors \cite{10.1109/JSEN.2010.2044239, 10.1109/JSEN.2019.2949435}, infrared cameras \cite{10.1186/1475-925X-10-93, 10.1109/EMBC.2016.7591489, 10.1109/ACCESS.2018.2845390}, and radar systems \cite{10.1109/JSEN.2020.3036039, 10.1109/JPROC.2023.3244362}. Among them, radar-based sensing offers unique advantages, including the ability to operate under low-light conditions and preserve privacy, thereby allowing continuous monitoring in diverse environments. In addition, microwaves and millimeter waves can penetrate bedding and clothing, allowing physiological signals to be measured through these materials.

Despite these advantages, a major challenge of radar-based physiological monitoring lies in the instability of its measurement accuracy. Previous studies have shown that accuracy can deteriorate depending on factors such as distance to the target and body orientation, underscoring the importance of radar placement \cite{10.1109/TMTT.2006.884652, 10.1109/TMTT.2013.2252186, 10.1109/APS.2016.7696293, 10.1109/JETCAS.2018.2811339, 10.1109/TIM.2023.3267348, 10.1109/ACCESS.2024.3434952, 10.1109/TMC.2024.3450318}. This degradation is believed to result from interference among multiple echoes reflected from different scattering centers on the human body. However, the precise mechanisms behind the accuracy deterioration remain unclear owing to the complex shape and motion of the human body. Therefore, developing a realistic radar simulation model is essential to analyzing and understanding the fundamental causes of accuracy degradation. 

The importance of simulators is further emphasized by recent developments in data-driven approaches. For instance, deep learning techniques have been increasingly applied to radar-based respiratory and cardiovascular monitoring \cite{10.1109/TMC.2022.3214721, 10.1109/JSEN.2023.3333025, 10.1109/TMC.2025.3563945}, requiring large-scale training datasets. Moreover, benchmarking signal processing algorithms on open datasets has become a common practice \cite{10.1109/RadarConf2147009.2021.9455239, 10.1109/MAES.2022.3216262, 10.1109/JBHI.2023.3240895}. Nonetheless, publicly available radar datasets involving physiological measurements remain scarce, primarily due to ethical and privacy concerns \cite{10.1038/s41597-020-00629-5}. In this context, simulation-based data augmentation offers a practical and ethical solution to address the scarcity of available datasets.

\begin{figure}[tb]
  \centering
  \includegraphics[width = 0.8\linewidth,pagebox=cropbox,clip]{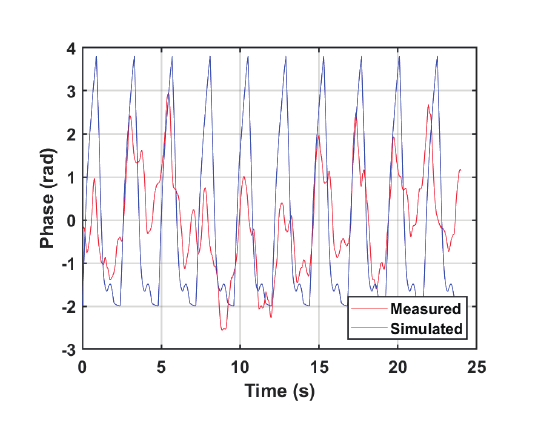}
  \caption{Results of simulated and measured radar displacement in \copyright2024 IEEE \cite{10.1109/RadarConf2458775.2024.10548317}.}
  \label{fig:conv_mukherjee}
\end{figure}

To address these challenges, we developed a radar simulator that models localized body surface displacements depending on radar positions, enabling a detailed analysis of echo interference. Conventional radar simulators for human targets typically focus on micro-Doppler signal generation for activity recognition, assuming anatomical joints as reflection sources and using motion capture data. However, few studies have simulated physiological signals such as respiration and heartbeat signals \cite{10.1109/RADAR.2008.4720816, 10.1109/MAES.2015.7119820, 10.1049/iet-rsn.2015.0065, 10.1109/RADAR.2018.8378798, 10.1109/MAES.2021.3138948, 10.1109/JSEN.2024.3386221, 10.1109/TRS.2024.3519138}. Most existing approaches model the human body using cylindrical representations or human phantoms and assume idealized, model-based displacements such as sinusoidal motions, rather than realistic geometries and displacements. For instance, Flouladi et al. used a single point chest reflection model with model-based displacements to simulate continuous wave radar signals \cite{10.1109/TSP.2013.6613953}. Hu et al. incorporated measured displacement waveforms into a cylindrical model to analyze multiple effects \cite{10.1109/TIM.2024.3420356}. Nahar et al. simulated stepped-frequency continuous-wave radar signals using a sinusoidally displaced cylindrical model with the method of moments \cite{10.1109/JETCAS.2018.2811339}. Mukherjee et al. used a human phantom with model-based displacements to investigate time-varying radar echoes \cite{10.1109/RadarConf2458775.2024.10548317}. However, despite the contributions of these traditional approaches, their accuracy remains limited. Figure \ref{fig:conv_mukherjee} compares the results of simulated and measured radar displacement as reported in \cite{10.1109/RadarConf2458775.2024.10548317}. The results reveal a notable mismatch between simulation and measurements. This discrepancy makes it challenging to thoroughly analyze the mechanisms of accuracy degradation. One possible reason for the inadequate accuracy in previous research is the reliance on model-based human geometry and displacements. This approach struggles to accurately simulate highly individual human body geometries as well as their complex deformations and displacements. Therefore, it is essential to employ realistic geometry and displacements instead of model-based ones.

Our previous work identified dominant reflection regions using real human body geometries captured by depth cameras \cite{10.1109/APUSNCURSINRSM.2018.8608643, 10.23919/APMC.2018.8617221, 10.1109/JSEN.2024.3519571}. However, existing simulators do not incorporate actual body displacements when simulating radar echoes. In this study, we propose a simulation framework that uses realistic human body geometry and displacement data acquired by a depth camera. This enables a rigorous comparison between simulated and measured radar signals. The main contributions of this work are as follows:
\begin{itemize}
  \item A simulation method for frequency-modulated continuous wave (FMCW) radar signals that uses human body geometry and displacement data acquired from a depth camera, replacing conventional simplified models.
  \item An experimental setup combining simultaneous radar measurements and depth camera recordings for model validation.
    \item Identification of dominant scattering regions and validation of Doppler velocity consistency between simulation data and measurements.
\end{itemize}

\section{Radar Signal and Electromagnetic Scattering}
\subsection{Radar Signal Model}
To simulate the signal received by an FMCW radar, it is essential to first define the transmitted and recorded signals. The transmitted signal $x^{\mathrm{T}}(\tau)$ is expressed as
\begin{equation}
  x^{\mathrm{T}}(\tau) = A\exp{\left\{\jj(2\pi f_0 \tau + \pi \gamma \tau^2)\right\}},
\end{equation}
where $A$ is the amplitude, $f_0$ is the starting frequency of the transmission, and $\tau$ represents the fast time. The chirp rate is defined as $\gamma = B/T_{\mathrm{c}}$, where $B$ is the sweep bandwidth and $T_{\mathrm{c}}$ is the sweep duration. 

Suppose that an FMCW radar system is equipped with $M$ virtual array antennas to measure $N$ point targets. The intermediate-frequency (IF) signal $x(\tau, t)$ is obtained from the filtered output of a mixer when feeding the transmitted and received signals to the mixer.
The IF signal $x_m(\tau, t)$ at the position $\bm{p}_m$ of the $m$th virtual array antenna is then expressed as
\begin{equation}
    \begin{split}
  x_{m}(\tau, t) = \sum_{n}^{N}A_n(t) \exp\left(  \jj\frac{4\pi\gamma R_{m,n}(t)}{c}\tau \right) \\
  \cdot\exp\left(  \jj \frac{4\pi f_0 R_{m,n}(t)}{c} \right),
    \end{split}
    \label{eq:sec2_distance}
\end{equation}
where $t$ is the slow time, $c$ is the speed of light, $A_n(t)$ is the amplitude depending on the echo intensity of the $n$th point target, and $R_{m,n}(t) = \|\bm{p}_m - \bm{r}^{\mathrm{P}}_n(t)\|$ is the distance between the position $\bm{p}_m$ of the $m$th antenna element and the position $\bm{r}^{\mathrm{P}}_n(t)$ of the $n$th point target. The effect of the antenna position on the amplitude $A_n(t)$ is assumed to be negligible due to the small aperture of the array antenna relative to $R_{m,n} (t)$.

\subsection{Visualization of Electromagnetic Scattering}
In a high-frequency approximation, electromagnetic scattering is often modeled by a finite number of equivalent radiating sources, which are referred to as scattering centers \cite{10.1109/TAP.1987.1144210, 10.1109/8.467641}. In this work, the scattering centers are estimated using the physical optics (PO) approximation \cite{10.1109/APUSNCURSINRSM.2018.8608643, 10.23919/APMC.2018.8617221, 10.1109/JSEN.2024.3519571}.
Let the human body region be represented by $V(t) \subseteq \mathbb{R}^3$ and the boundary surface of the region $V(t)$ be denoted by $\partial V(t)$.
First, assuming that the target is a perfect electric conductor, the surface current density $\bm{K}(\bm{r})$ induced at $\bm{r} \in \partial V(t)$ on the target surface is calculated using the PO approximation as
\begin{align}
  \bm{K}(\bm{r})&=2\bm{n}(\bm{r})\times\bm{H}^{\mathrm{i}}(\bm{r}),
  \label{eq:POscattering}
\end{align}
where $\bm{n}(\bm{r})$ is the outward normal vector of $\partial V(t)$ at $\bm{r}$ and $\bm{H}^{\mathrm{i}}(\bm{r})$ represents the incident magnetic field. 

Assuming vertical polarization along the $z$-axis for both the transmitting and receiving antennas, the $z$-factor of the electric field $E_z(\bar{\bm{p}}, t)$ at an observation position $\bar{\bm{p}}$ can be calculated using the induced surface current density $\bm{K}(\bm{r})$ on the scatterer surface:
\begin{align}
  E_z(\bar{\bm{p}}, t)           & = -\mathrm{j}\omega\mu \iint_{\partial V(t)} \bm{G}(\bar{\bm{p}}, \bm{r})\cdot\bm{K}(\bm{r}) \, \dd S, \label{eq:dyadicEwave}\\
  \bm{G}(\bar{\bm{p}}, \bm{r}) & = D_{\mathrm{R}}(\theta, \phi)\frac{\mathrm{e}^{-\mathrm{j}k \|\bar{\bm{p}} - \bm{r}\|}}{4\pi \|\bar{\bm{p}} - \bm{r}\|}\hat{\bm{z}}\cdot\left( \bar{\bar{I}}+ \frac{1}{k^2}\nabla\nabla\right),
\end{align}
where $\bar{\bm{p}} = (1/M) \sum_{m} \bm{p}_m$ is the centroid of the positions of the $M$ antennas, $\omega$ is the angular frequency, $\mu$ is the permeability, $\bm{G}(\bar{\bm{p}}, \bm{r})$ is the Green function, and $\dd S$ is the surface element. $D_{\mathrm{R}}(\theta, \phi)$ is the directivity pattern of the receive antenna, $\theta$ is the azimuth angle, $\phi$ is the elevation angle, and $\hat{\bm{z}}$ is the unit vector in the direction of the $z$ axis.
$\bar{\bar{I}}$ is the unit dyad, $k$ is the wavenumber, and $\nabla\nabla$ denotes the dyadic operator.

The scattering centers are visualized using a weighting function $w_{\mathrm{eye}}$ as a kernel function of an integral \cite{1570291227534485888}:
\begin{align} 
  P^{\mathrm{S}}(\bar{\bm{p}}, \bm{r}) 
  &= \bigg| -\mathrm{j}\omega\mu 
    \iint_{\partial V(t)} w_{\mathrm{eye}}( \bm{r}, \bm{\rho} ) \bm{G}(\bar{\bm{p}}, \bm{\rho}) \cdot \bm{K}(\bm{\rho})  \, \dd S \bigg|^2, \\
   w_{\mathrm{eye}}(\bm{r}, \bm{\rho}) &=
  \begin{cases}
    \displaystyle\frac{1}{2}\left\{\cos\left(\frac{\pi\|\bm{r}-\bm{\rho}\|}{a_0}\right)+1\right\} & \|\bm{r}-\bm{\rho}\| < a_0 \\
    0                                                                                 & \text{otherwise,}
  \end{cases}
  \label{eq:PO_eyfunction}
\end{align}
where $\bm{\rho} \in \partial V(t)$ is a point in the human body geometry, and $a_0$ is the radius of the local integration region.

$P^{\mathrm{S}}(\bar{\bm{p}}, \bm{r})$ represents the contribution to the power of the electromagnetic wave observed at position $\bar{\bm{p}}$ from the local area around the point $\bm{r}$. In this study, a finite number of scattering centers are extracted from this power distribution, and they are used as the finite set of targets in Eq. \eqref{eq:sec2_distance} to generate the simulated radar signal. In addition, this power is used to determine the amplitude values $A_n(t)$ of the different targets. The detailed procedure for generating the simulated signal is presented in the following section.

\section{Proposed Simulation Method}
\subsection{Radar Echo Simulation Based on Scattering Centers}
This section explains a method for simulating radar echoes from a human body geometry. First, the local maxima of the power $P^{\mathrm{S}}(\bar{\bm{p}}, \bm{r})$ are obtained using the method described in the previous section, and the scattering centers are identified. The set of scattering centers is defined as
\begin{align}
  \mathcal{Q}(t) &= \bigg\{ \bm{r} \in  \partial V(t) \:\bigg|\: \nabla P^{\mathrm{S}}(\bar{\bm{p}}, \bm{r}) = \bm{0}, \nonumber 
  \\ & \qquad \nabla\nabla^{\mathrm{T}} P^{\mathrm{S}}(\bar{\bm{p}},  \bm{r}) < 0, P^{\mathrm{S}}(\bar{\bm{p}}, \bm{r}) > \theta_{P^{\mathrm{S}}} \bigg\},
  \label{eq:sec3ExtractSC}
\end{align}
where $\nabla$ is the gradient with respect to $\bm{r}$, and $\nabla\nabla^\mathrm{T} P^{\mathrm{S}}(\bar{\bm{p}}, \bm{r}) < 0$ indicates that the Hessian matrix is negative definite. $\theta_{P^{\mathrm{S}}}$ is a threshold parameter used to detect the scattering center while avoiding detecting noise.

Let $\mathcal{Q}(t)$ be a set of $N_{\mathcal{Q}}(t)=|\mathcal{Q}(t)|$ elements at time $t$. The amplitude $A_n(t)$ and distance $R_{m,n}(t)$ in Eq. \eqref{eq:sec2_distance} are then given using the scattering center position $\bm{q}_n \in \mathcal{Q}(t)$ ($n=1, 2, \cdots, N_{\mathcal{Q}}(t)$) and antenna element position $\bm{p}_m$ ($m=1, 2, \cdots, M$):
\begin{align}
  A_n(t) &= \sqrt{P^{\mathrm{S}}(\bar{\bm{p}}, \bm{q}_n)} \eta_n(t), \\
  R_{m,n}(t) &=\| \bm{p}_m - \bm{q}_n \|,
\end{align}
where $\eta_n(t)$ is the complex phase term for scattering, taking into account factors such as the electromagnetic properties of the scatterer and the angle of arrival, which are used to simulate radar echoes from a target human body with respiratory motion. This study uses scattering centers derived from time series of the human body geometry obtained via a depth camera to enable the generation of simulation signals for physiological signal measurement, without relying on model-based body geometry or displacement.

\subsection{Reduction of the Computational Load in Simulation}
The method described in the previous section requires repeated computation of Eqs. \eqref{eq:dyadicEwave}--\eqref{eq:PO_eyfunction} at each time step, resulting in a prohibitive computational load. Furthermore, the depth-camera images contain measurement noise and random fluctuations. To mitigate these issues, we employ a time-averaged human body geometry instead of using raw depth data. To compute the time-averaged geometry, we first estimate the depth image
$z_{\mathrm{c}}(u_{\mathrm{x}},u_{\mathrm{y}},t)$ from the body surface $\partial V(t)$, where $(u_{\mathrm{x}},u_{\mathrm{y}})$ denotes the image coordinates of the depth camera. We then average $z_{\mathrm{c}}$ over time to obtain
$\bar{z}_{\mathrm{c}}(u_{\mathrm{x}},u_{\mathrm{y}}) = \frac{1}{T}\int_{0}^{T} z_{\mathrm{c}}(u_{\mathrm{x}},u_{\mathrm{y}},t) \dd t$. The resulting image $\bar{z}_{\mathrm{c}}$ is used to reconstruct the time-averaged human body geometry $\overline{\partial V}$.

The scattering centers $\bar{\bm{q}}_n \in \bar{\mathcal{Q}}$ are then extracted from the geometry averaged over time $\overline{\partial V}$ following Eq. \eqref{eq:sec3ExtractSC}, where $A_n$ is estimated as in Eq.~(9) while $R_{m,n}(t)$ is estimated as
\begin{align}
  R_{m,n}(t) &=\bigg\|\bm{p}_m - \underset{\bm{r} \in  \partial V(t)} {\textrm{argmax}}\:(\bm{p}_m - \bm{r}) \cdot ( \bm{p}_m- \bar{\bm{q}}_n) \bigg\|.
  \label{eq:C3SCDistanceSimple}
\end{align}
 Applying the procedures in Eqs.~\eqref{eq:dyadicEwave}--\eqref{eq:PO_eyfunction} at each time step to calculate the distances of the scattering centers is computationally expensive, and a simplified method based on Eq.~\eqref{eq:C3SCDistanceSimple} is thus employed.

\begin{figure}[tb]
  \centering
  \begin{minipage}[c]{0.55\linewidth}
    \centering
    \subfloat[]{
    \includegraphics[width=\linewidth,pagebox=cropbox,clip]{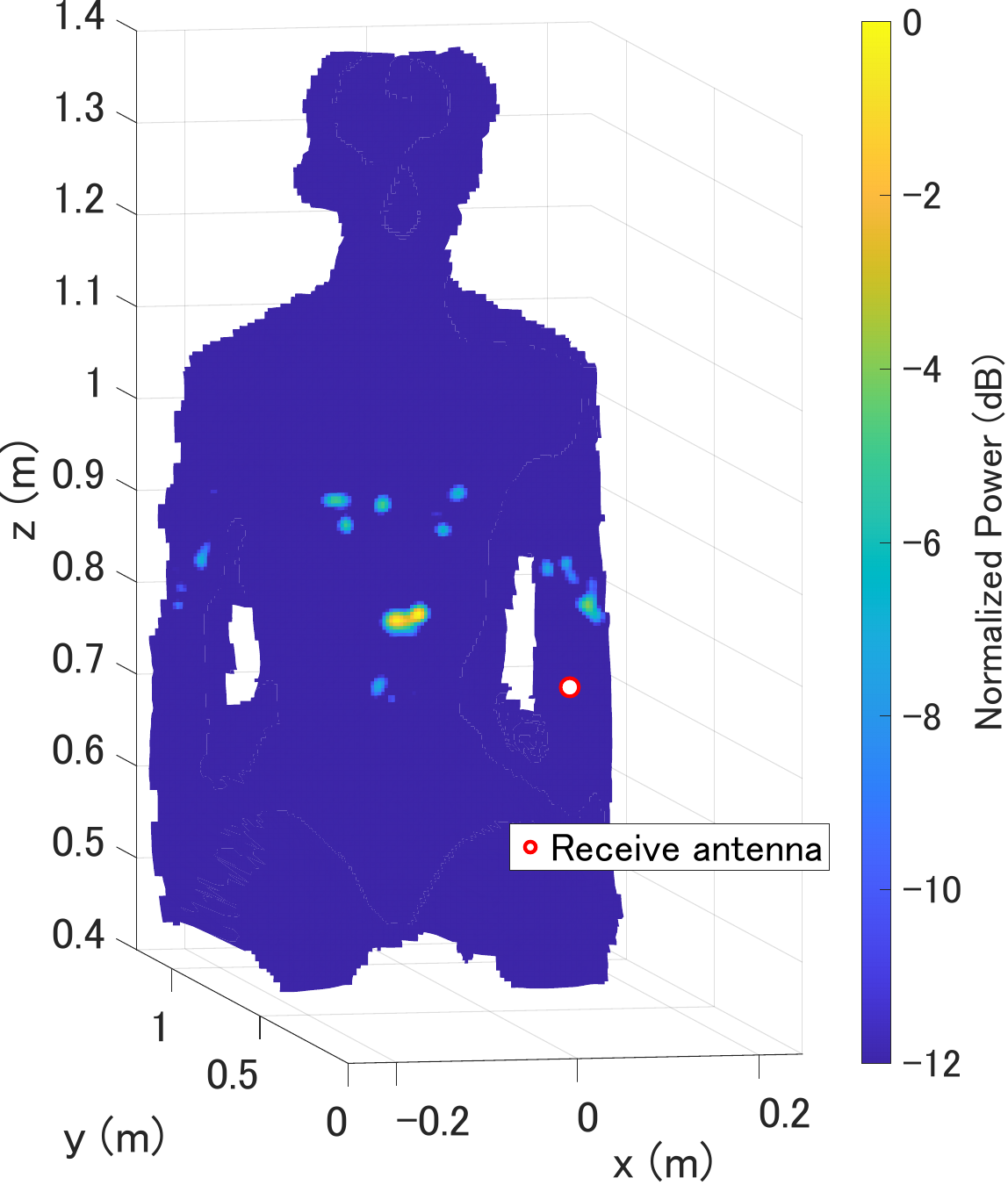}
     \label{fig:scatteringCenterFront_PO}
      }
  \end{minipage}
  \hfill
  \begin{minipage}[c]{0.4\linewidth}
    \centering
    \subfloat[]{
    \includegraphics[width=\linewidth,pagebox=cropbox,clip]{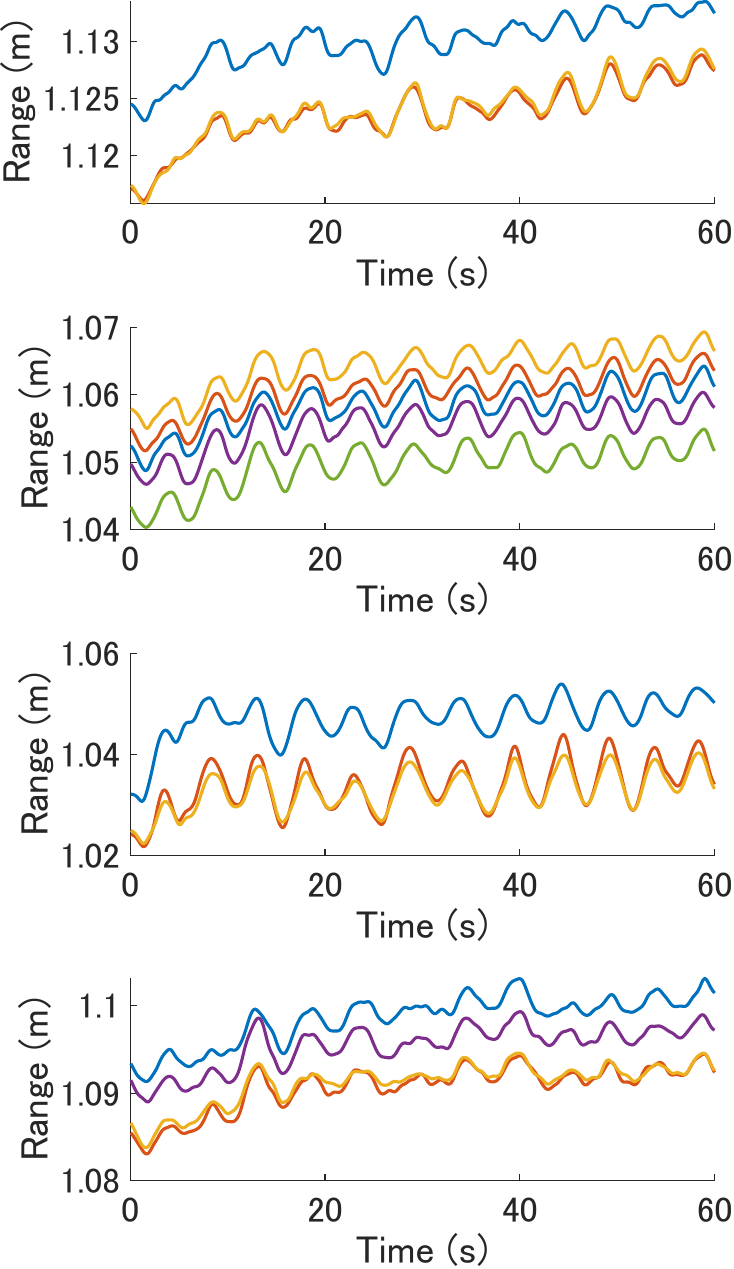}
       \label{fig:scatteringCenterFront_distance}
    }
  \end{minipage}
  \caption{Scattering power (a) and distances to the scattering centers (b) at the front of the human.}
    \label{fig:scatteringCenterFront}
\end{figure}

We present examples of $P^{\mathrm{S}}(\bar{\bm{p}}, \bm{r})$ and $R_{m,n}$ for the front side of the human torso (where the condition ID corresponds to C1 in Table \ref{tbl:expCondition}). Detailed experimental conditions are provided in Section \ref{sec:EXPRESULTS}. Figure~\ref{fig:scatteringCenterFront_PO} shows the scattering power $P^{\mathrm{S}}(\bar{\bm{p}}, \bm{r})$ obtained when measuring the front side of the human body using a millimeter-wave radar system. Strong reflections are observed in the arms, chest, and abdomen. Figure~\ref{fig:scatteringCenterFront_distance} shows the distance $R_{m,n}(t)$ from the scattering centers with a scattering power higher than $-10$ dB relative to the maximum. Three, five, three, and four scattering centers can be extracted for the right arm, chest, abdomen, and left arm, respectively. Figure \ref{fig:scatteringCenterFront_distance} indicates that the scattering centers within each body part exhibit highly similar motion. Furthermore, while correlated components of respiratory motion were observed in different regions, movements were not entirely identical. These results demonstrate that the proposed method can simulate radar-received signals by superimposing multiple echoes, using the corresponding amplitudes and distances described above.

\section{EXPERIMENTAL ACCURACY EVALUATION}\label{sec:EXPRESULTS}
\subsection{Experimental Setup}
\begin{figure}[tb]
  \centering
  \includegraphics[width = 1.0\linewidth,pagebox=cropbox,clip]{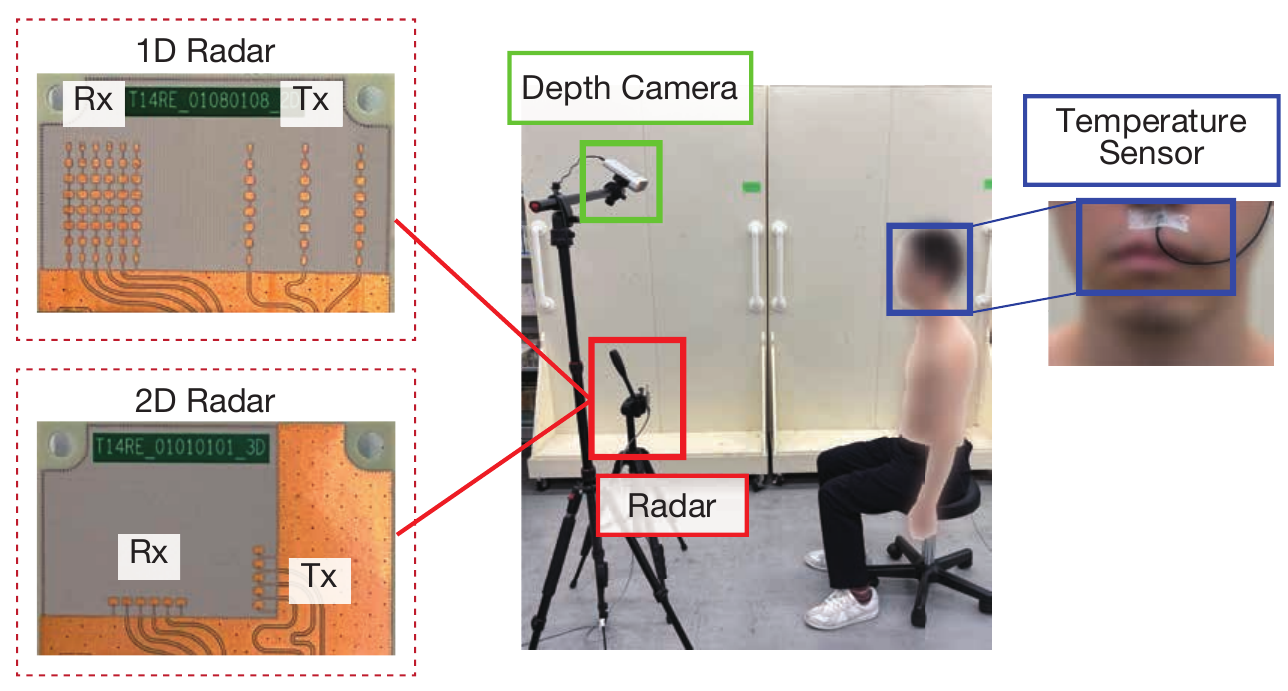}
  \caption{Experimental setup.}
  \label{fig:expsetup}
\end{figure}

\begin{figure}[tb]
  \centering
  \subfloat[]{
    \includegraphics[width=0.46\linewidth,pagebox=cropbox,clip]{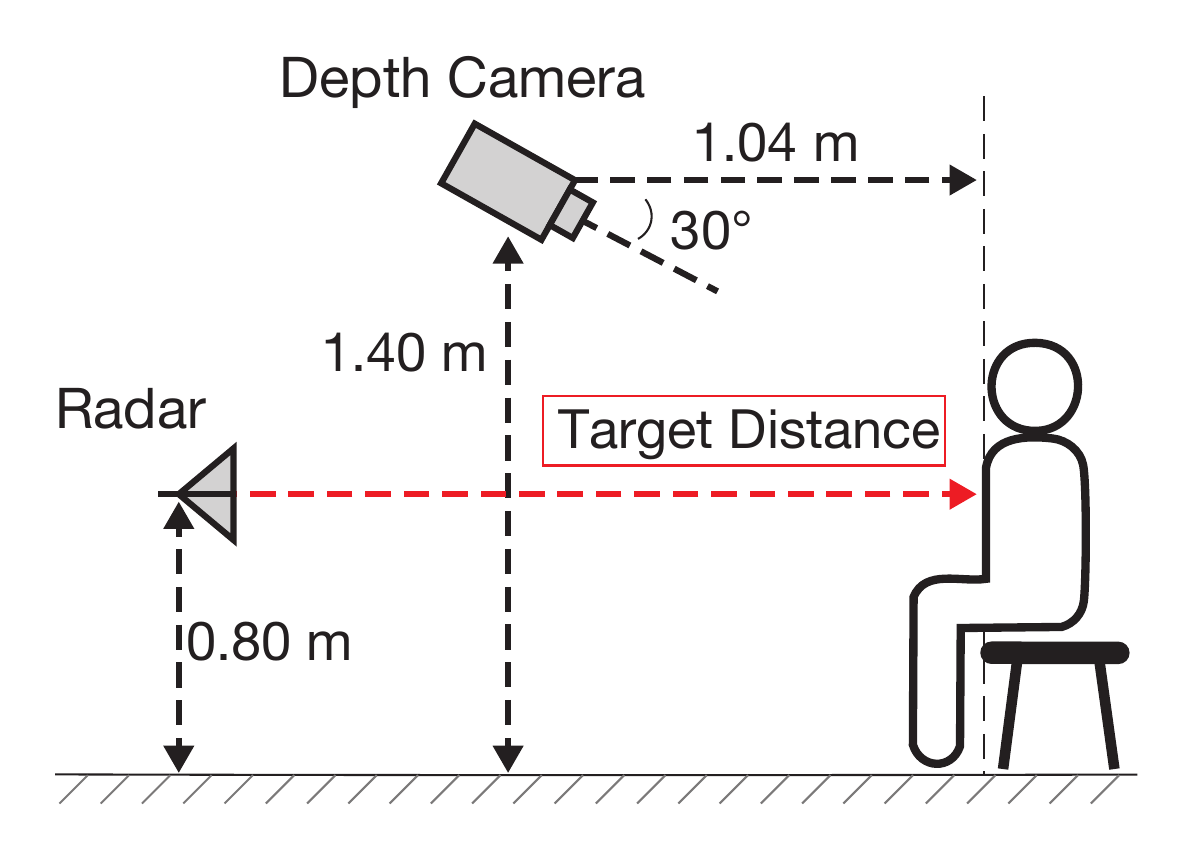}
  }
  \hfill
  \subfloat[]{
    \includegraphics[width=0.46\linewidth,pagebox=cropbox,clip]{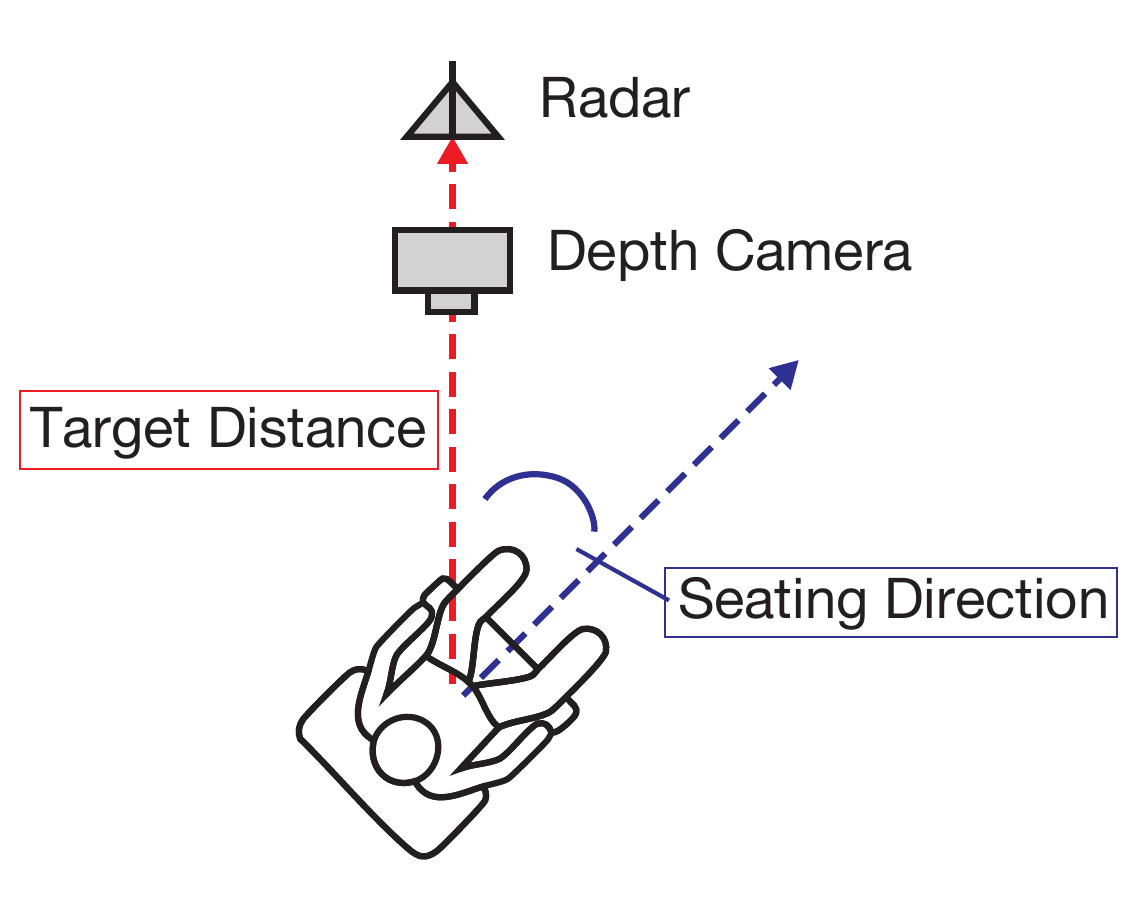}
  }
  \caption{Experimental setup from (a) the side view and (b) the top view.}
  \label{fig:experimentalDiagram}
\end{figure}

A simultaneous measurement experiment was conducted using a radar system and a depth camera. Specifically, we used two types of radar device with a 1D array and 2D array (T14RE\_01080108\_2D and T14RE\_01010101\_3D; S-Takaya Electronics Industry, Okayama, Japan), whose specifications are listed in Table~\ref{tbl:radar_system}. In addition, we used an Azure Kinect DK depth camera (Microsoft Corp., Redmond, WA, USA), whose specifications are listed in Table~\ref{tbl:depth_camera}. The experimental setup is shown in Figs.~\ref{fig:expsetup} and \ref{fig:experimentalDiagram}. The target distance ranged from 1.0 to 3.0~m, and the seating direction varied from \ang{0} to \ang{180}, as summarized in Table~\ref{tbl:expCondition}. For an accurate evaluation, reference respiratory data were recorded using a nasal temperature sensor (AP-C050) with a biomedical signal measurement system (Polymate Pro MP6000; Miyuki Giken, Tokyo, Japan). Six male participants, having a mean age of 22.3 years (SD $\pm 0.7$ years), took part in the experiment. The duration of each measurement was $T=60.0$~s.

\begin{table}[tb]
  \centering
  \caption{EXPERIMENTAL CONDITIONS}
  \resizebox{0.8\linewidth}{!}{
    \setlength{\tabcolsep}{0.2em}
    \begin{tabular}{c|ccc}
      \toprule
      Condition ID & Target Distance      & Seating Direction        & Radar Type          \\  \midrule
      C1           & 1.0 m                  & \multirow{3}{*}{\ang{0}} & \multirow{3}{*}{1D} \\
      C2           & 2.0 m                  &                          &                     \\
      C3           & 3.0 m                  &                          &                     \\\midrule
      C4           & \multirow{4}{*}{1.0 m} & \ang{45}                 & \multirow{4}{*}{1D} \\
      C5           &                      & \ang{90}                 &                     \\
      C6           &                      & \ang{135}                &                     \\
      C7           &                      & \ang{180}                &                     \\\midrule
      C8           & 2.0 m                  & \ang{0}                  & 2D                  \\
      \bottomrule
    \end{tabular}
  }
  \label{tbl:expCondition}
\end{table}

\begin{table}[tb]
  \centering
  \caption{SPECIFICATIONS OF THE RADAR SYSTEM}
  \begin{threeparttable}[h]
    \begin{tabular}{l|cc}
      \toprule
      Radar Type                   & 1D                                    & 2D                                     \\
      \midrule
      Central frequency             & \multicolumn{2}{c}{79 GHz}                                                     \\
      Bandwidth                    & \multicolumn{2}{c}{3.354 GHz}                                                  \\
      Num. of Tx                   & \multicolumn{2}{c}{3}                                                          \\
      Tx element spacing           & 7.6 mm                                & 1.9 mm                                 \\
      Output power (EIRP)\tnote{*} & 24 dBm                                & 16 dBm                                 \\
      Tx element beamwidth         & \multirow{2}{*}{$\pm$35 / $\pm$4 deg} & \multirow{2}{*}{$\pm$33 / $\pm$45 deg} \\
      (azimuth/elevation)          &                                       &                                        \\
      Num. of Rx                   & \multicolumn{2}{c}{4}                                                          \\
      Rx element spacing           & 1.9 mm                                & 1.9 mm                                 \\
      Rx element beamwidth         & \multirow{2}{*}{$\pm$45 / $\pm$4 deg} & \multirow{2}{*}{$\pm$45 / $\pm$45 deg} \\
      (azimuth/elevation)          &                                       &                                        \\
      Sampling frequency           & \multicolumn{2}{c}{100 Hz}                                                     \\
      \bottomrule
    \end{tabular}
    \begin{tablenotes}
      \item[*]EIRP: Equivalent Isotropically Radiated Powers
    \end{tablenotes}
  \end{threeparttable}
  \label{tbl:radar_system}
\end{table}

\begin{table}[tb]
  \centering
  \caption{SPECIFICATIONS OF THE DEPTH CAMERA}
  \begin{tabular}{l|c}
    \toprule
                     & Specification        \\
    \midrule
    Lidar method     & ToF (time of flight) \\
    Camera pixels    & $512\times512$              \\
    Field of view    & $120^{\circ} \times 120^{\circ}$       \\
    Sampling freq.   & 15 Hz                \\
    Operating range  & 0.25 m $\sim$ 2.88 m      \\
    Depth resolution & 1 mm                 \\
    \bottomrule
  \end{tabular}
  \label{tbl:depth_camera}
\end{table}

\subsection{Parameters of the Proposed Method}
In the simulation, white Gaussian noise was introduced by superimposing a zero-mean, normally distributed complex random sequence on the IF time series to achieve a signal-to-noise ratio of $-20$ dB. The threshold for the scattering power was set at $\theta_{P^{\mathrm{S}}}=-20$ dB. The complex phase term of scattering $\eta_n(t)$ was set to the opposed phase. The parameters used for the electromagnetic scattering simulation were consistent with those used in our previous studies~\cite{10.1109/APUSNCURSINRSM.2018.8608643, 10.1109/JSEN.2024.3519571, 10.23919/APMC.2018.8617221}. The angular frequency of the electromagnetic wave is given by $\omega=2\pi f_{\mathrm{c}}$, where $ f_{\mathrm{c}}=79$~GHz denotes the central frequency of the FMCW radar. The local integration radius was defined as $a_0 = 5\lambda=5c/f_{\mathrm{c}}$. The directivity of the antenna for both transmission and reception was based on the specifications listed in Table \ref{tbl:radar_system}.

\subsection{Evaluation Metrics}\label{sec:EVALUATION}
To assess the fidelity of the simulated IF signal to the actual measurements, we examine three aspects: the radar image, the phase at the point of maximum power, and the spectrogram.

\subsubsection{Radar Image}
A radar image $I(r,\theta,\phi,t)$ is obtained as
\begin{align}
  I(r,\theta,\phi,t) = \sum_{m=1}^{M}w_m^{\mathrm{A}} a^*_m(\theta,\phi) y_m(r,t), 
\end{align}
where $r$ is the range, $w_m^{\mathrm{A}}$ is the window function (Taylor window for 1D arrays and a rectangular window for 2D arrays), and $a^*_m(\theta,\phi)$ is the complex conjugate of the steering vector. $y_m(r,t)$ is the signal received using the $m$th antenna element, expressed as
\begin{align}
  y_{m}(r,t) =  \int_{0}^{T_{\mathrm{c}}} w^{\mathrm{R}}(\tau) x_{m}(\tau, t)\exp\left(-\jj2\pi  \frac{2Br}{cT_{\mathrm{c}}}\tau\right) \: \dd\tau,
\end{align}
where $w^{\mathrm{R}}(\tau)$ is the window function (Taylor window). The elevation angle $\phi$ is used only for 2D arrays, and the same applies in the following discussion.

Stationary components are not of interest in this analysis, and the time-averaged component is thus subtracted from the radar image:
\begin{align}
  \tilde{I}(r,\theta,\phi,t)  = I(r,\theta,\phi,t) - \frac{1}{T}\int_0^{T} I(r,\theta,\phi,t) \, \dd t.
\end{align}
Its power is computed as
\begin{align}
  I_{\mathrm{A}}(r,\theta,\phi)   & = \frac{1}{T}\int_0^{T}  \left|\tilde{I}(r,\theta,\phi,t)\right|^2 \dd t,
\end{align}
where $T$ is the measurement period. To evaluate the similarity between the measured and simulated radar images $I_{\mathrm{A}}(r,\theta,\phi)$ and $\hat{I}_{\mathrm{A}}(r,\theta,\phi)$, we evaluate the maximum value of a cross-correlation function given by
\begin{equation}
  \begin{split}
    \rho_{\mathrm{I}} = \max_{\varDelta r,\varDelta\theta, \varDelta\phi} & \frac{1}{\Gamma_{\mathrm{A}} \hat{\Gamma}_{\mathrm{A}}} \iiint_{-\infty}^{\infty} \chi(r,\theta,\phi) I_{\mathrm{A}}(r,\theta,\phi) \\
                                                                          & \cdot \hat{I}_{\mathrm{A}}(r+\varDelta{r},\theta+\varDelta\theta,\phi+\varDelta\phi) \; \dd r \dd\theta \dd\phi,
  \end{split}
\end{equation}
\begin{equation}
  \chi(r,\theta,\phi) =
  \begin{cases}
    1 & \mathrm{if} \, r\in\mathcal{E}_{\mathrm{r}} \land \theta\in\mathcal{E}_{\theta} \land  \phi\in\mathcal{E}_{\phi} \\
    0 & \mathrm{otherwise},
  \end{cases}
\end{equation}
where $\Gamma_{\cdot}$ is a normalization constant, and $\chi(r,\theta,\phi)$ is an indicator function that specifies the target evaluation region.
$\mathcal{E}_{r}$, $\mathcal{E}_{\theta}$, and $\mathcal{E}_{\phi}$ represent the evaluation ranges. The evaluation range is defined as $\mathcal{E}_{r} = [r^* - r_0, r^* + r_0]$ and $\mathcal{E}_{\theta} = \mathcal{E}_{\phi} = [-\ang{45}, \ang{45}]$, where $r^*$ represents the target range and $r_0=0.5$~m is empirically set.

\subsubsection{Displacement} The target displacement $d(t)$ is calculated as
\begin{align}
  d(t) = \frac{\lambda}{4\pi} \mathrm{unwrap}\left[\angle \tilde{I}(r_{\mathrm{max}}, \theta_{\mathrm{max}}, \phi_{\mathrm{max}},t)\right],
  \label{eq:phaseUnwrap}
\end{align}
where $(r_{\mathrm{max}}, \theta_{\mathrm{max}}, \phi_{\mathrm{max}})$ is determined by
\begin{align}
  (r_{\mathrm{max}}, \theta_{\mathrm{max}}, \phi_{\mathrm{max}}) = \underset{(r, \theta, \phi)}{\textrm{argmax}} \; I_{\mathrm{A}}(r,\theta,\phi).
  \label{eq:phaseExtractedRadarImage}
\end{align}
Given the measured displacement $d(t)$ and simulated displacement $\hat{d}(t)$, the correlation coefficient $\rho_{\mathrm{d}}$ and root mean square (RMS) error $\epsilon_{\mathrm{d}}$ are defined as
\begin{align}
  \rho_{\mathrm{d}}     & = \frac{1}{\Gamma_{\mathrm{d}} \hat{\Gamma}_{\mathrm{d}}}\int_{0}^{T} d(t) 
  \hat{d}(t) \, \dd t , \label{eq:correlation_d}\\
  \epsilon_{\mathrm{d}} & = \sqrt{\frac{1}{T} {\| d(t) - \hat{d}(t) \|}^2}.\label{eq:RMSE_d}
\end{align}

Typical respiratory displacements have a dominant component around 0.2 Hz, and we thus apply a high-pass filter to $d(t)$ to obtain the respiratory displacement $d_{\mathrm{HF}}(t)$, using a fifth-order Butterworth filter with a cutoff frequency of 0.05 Hz. Similar to Eqs. \eqref{eq:correlation_d} and \eqref{eq:RMSE_d}, the correlation coefficient $\rho^{\mathrm{HF}}_{\mathrm{d}}$ and the RMS error $\epsilon^{\mathrm{HF}}_{\mathrm{d}}$ are computed for the measured displacement $d_{\mathrm{HF}}(t)$ and simulated displacement $\hat{d}_{\mathrm{HF}}(t)$.

\subsubsection{Spectrogram} The spectrogram $X(f,t)$ is calculated as
\begin{align}
   X(f,t)                    = \left|\int_{-\infty}^{\infty}\tilde{x}_{m_0}(\tau_0,t')w(t'-t)\exp(-\jj2\pi f t') \, \dd t'\right|^2,
\end{align}
where $m_0$ and $\tau_0$ are the arbitrary array element index and fast time index, respectively; $w(t' - t)$ is a Taylor window function with a duration of 0.5 s, and $ \tilde{x}_{m_0}(\tau,t)$ is defined by
\begin{align}
  \tilde{x}_{m_0}(\tau,t) = \mathrm{BPF} \left[ x_{m_0}(\tau,t) - \bar{x}_{m_0}(\tau)\right] .
 \label{eq:Sec4BPFforSpectrogram}
\end{align}
Here, $\bar{x}_{m_0}(\tau)=(1/T) \int_{0}^{T} x_{m_0}(\tau,t) \dd t$ represents the static component of the IF signal. The term BPF in Eq. \eqref{eq:Sec4BPFforSpectrogram} refers to a bandpass filter that extracts velocities of targets within the range $r \in \mathcal{E}_{r} = [r^* - r_0, r^* + r_0]$. The coefficient of correlation $\rho_{\mathrm{S}}$ between the measured and simulated spectrograms is calculated for performance evaluation after converting the power values to a decibel scale, to account for variations in low-power components.

\subsection{Performance Evaluation of the Proposed Method}
\begin{figure}[tb]
  \centering
  \subfloat[\label{fig:C4_RadarImageSub1C1Sim}]{
    \includegraphics[width=0.45\linewidth,pagebox=cropbox,clip]{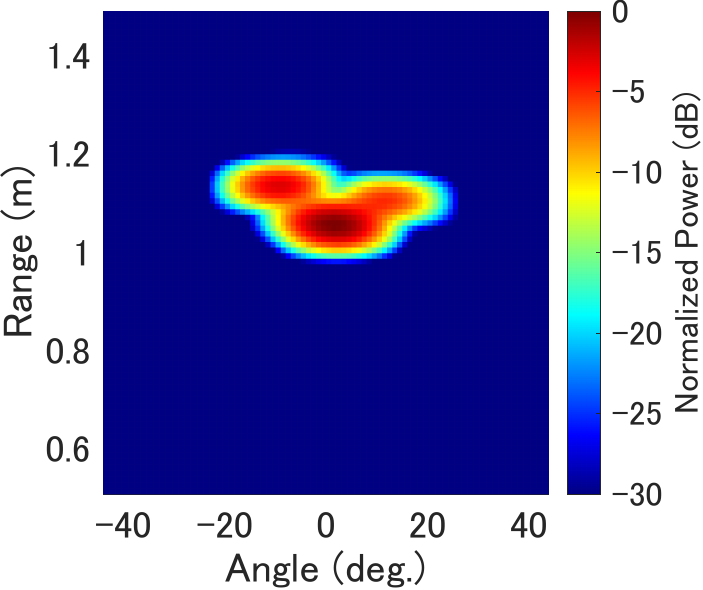}   
  }
  \hfill
  \subfloat[\label{fig:C4_RadarImageSub1C1Actual}]{
    \includegraphics[width=0.47\linewidth,pagebox=cropbox,clip]{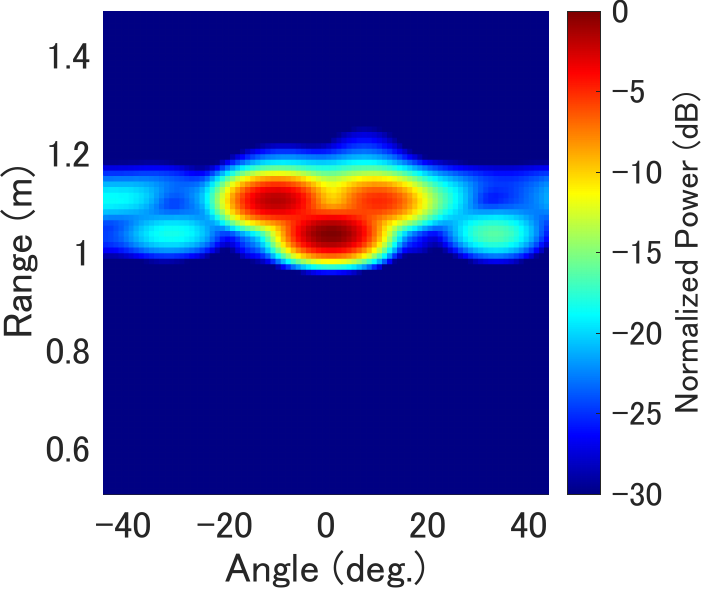}
  }
  \\
  \subfloat[]{
    \includegraphics[width=0.47\linewidth,pagebox=cropbox,clip]{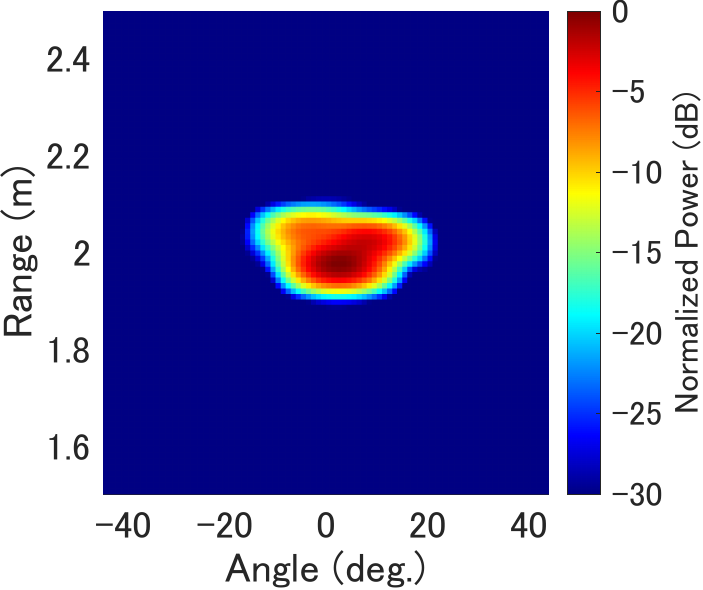}
  }
  \hfill
  \subfloat[]{
    \includegraphics[width=0.47\linewidth,pagebox=cropbox,clip]{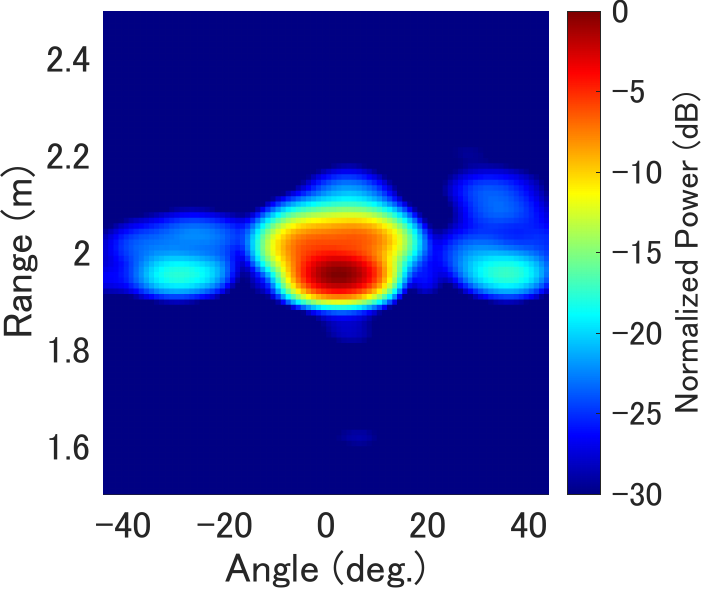}
  }
  \\
  \subfloat[]{
    \includegraphics[width=0.47\linewidth,pagebox=cropbox,clip]{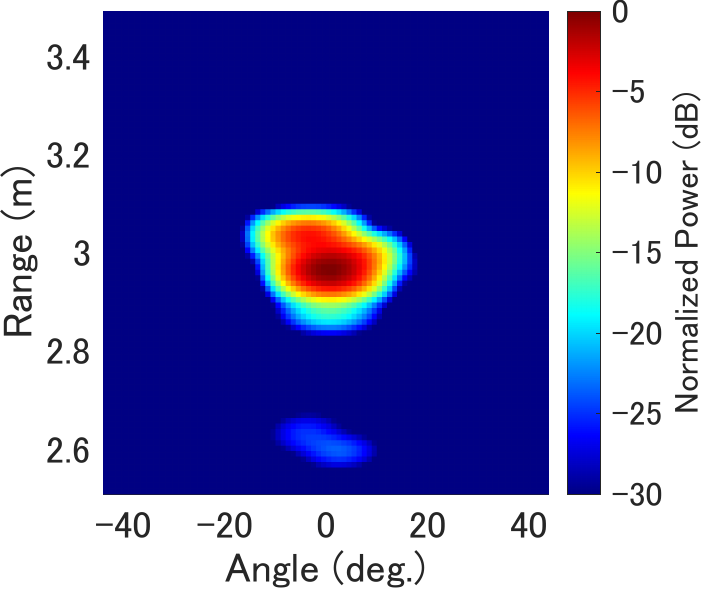}
  }
  \hfill
  \subfloat[]{
    \includegraphics[width=0.47\linewidth,pagebox=cropbox,clip]{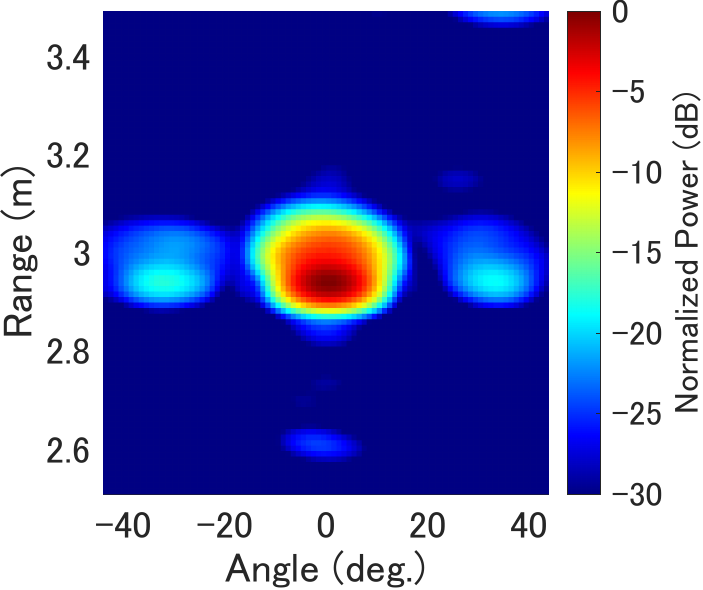}
  }
  \caption{Radar images obtained by simulation and measurement; (a) 1.0-m simulation, (b) 1.0-m measurement, (c) 2.0-m simulation, (d) 2.0-m measurement, (e) 3.0-m simulation, and (f) 3.0-m measurement.}
  \label{fig:C4_RadarImageDistance}
\end{figure}

Table~\ref{tbl:evalDistance} presents the average accuracy of the proposed method for all participants under each condition. First, the radar image correlation coefficient $\rho_{\mathrm{I}}$ was 0.938 on average across all conditions, demonstrating consistent performance. Figure~\ref{fig:C4_RadarImageDistance} shows simulated and measured radar images for participant 1 under conditions C1, C2, and C3. The correlation coefficients of the radar images for the conditions were 0.965, 0.966, and 0.988, respectively. Under C1, echoes from three parts of the body---the right arm, torso, and left arm---are observed in both the simulated and measured radar images. By contrast, under the more distant conditions of C2 and C3, it is observed that the three body parts are positioned closer to each other. This is because, as the distance increases, the relative angular separation between body parts as seen from the radar becomes smaller. 

\begin{figure}[tb]
  \centering
  \includegraphics[width = 0.8\linewidth,pagebox=cropbox,clip]{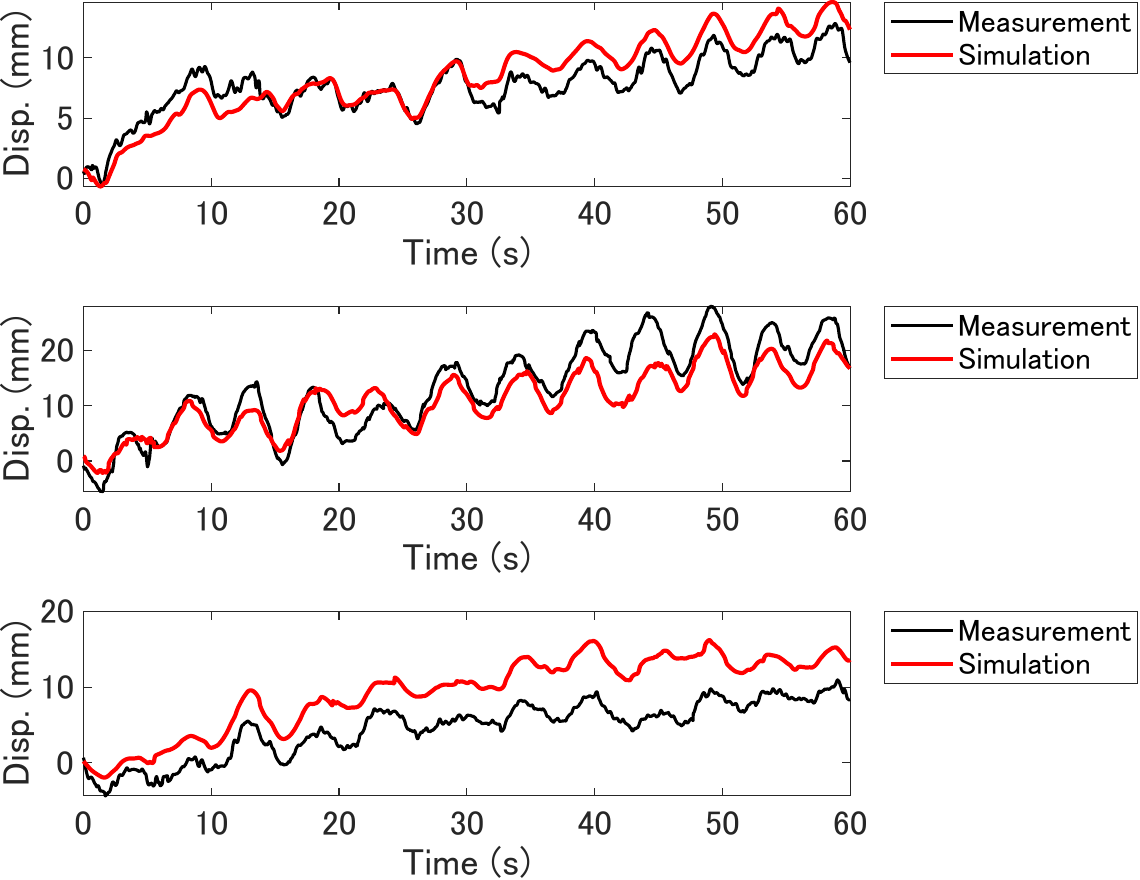}
  \caption{Radar displacements in simulation and measurements (top: right arm, middle: torso, bottom: left arm).}
  \label{fig:sub1PhaseDisp}
\end{figure}

The displacement correlation coefficient $\rho_{\mathrm{d}}$ was then found to be 0.562, with an RMS error of 4.52 ms. Under C1, particularly high accuracy was achieved, with an average correlation coefficient of 0.750. Figure~\ref{fig:sub1PhaseDisp} shows the displacements for the right arm, torso, and left arm of participant 1 under C1. These three body parts correspond to the peaks extracted from the radar images shown in Figs. \ref{fig:C4_RadarImageSub1C1Sim} and \ref{fig:C4_RadarImageSub1C1Actual}. The correlation coefficients for the three parts were 0.928, 0.936, and 0.964, respectively, confirming a strong correlation between the simulation and measurement. Although accuracy was high under near-range conditions, the performance degraded appreciably under far-range conditions (C3) and for the 2D radar condition (C8), with correlation coefficients of 0.208 and 0.461, respectively. This degradation arises from the increased mixing of reflections from multiple body parts at greater distances, making it difficult for simulations to replicate the demodulation of mixed IF signals in Eq.~\eqref{eq:phaseUnwrap}.

The correlation coefficient of the high-pass filtered displacement averaged 0.685 across all conditions, with the RMS error being 0.97~ms. Notably, high accuracy was achieved under C1, C2, and C4, with correlation coefficients of 0.803, 0.826, and 0.830, respectively, indicating the accurate reproduction of respiratory components. By contrast, in terms of RMS error, the conditions under which the radar faced the front of the participants (C1, C2, C3, and C8) showed higher errors than the other conditions. This is likely due to larger absolute displacements caused by respiration in front-facing orientations, increasing susceptibility to error.

\begin{figure}[tb]
  \centering
  \subfloat[]{
    \includegraphics[width=0.47\linewidth,pagebox=cropbox,clip]{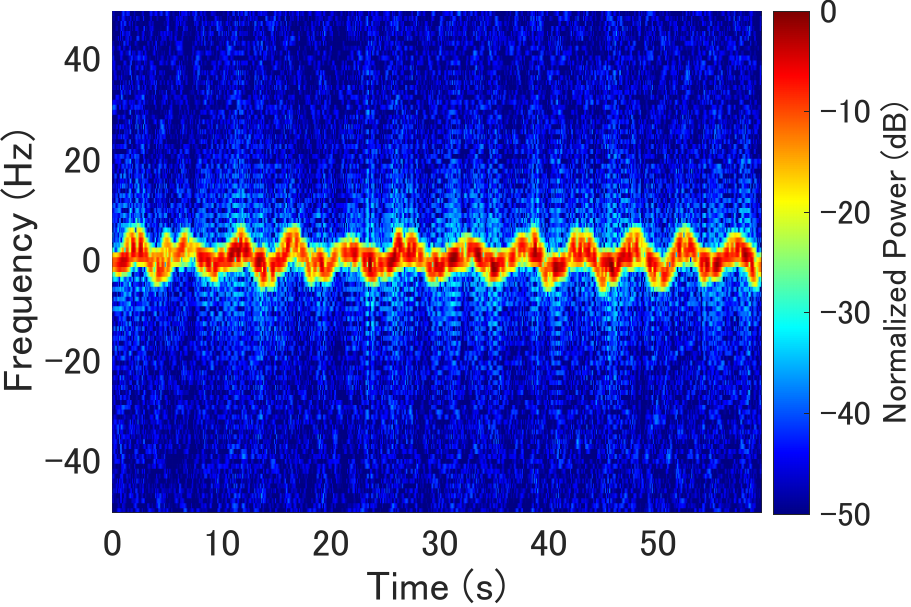}
  }
  \hfill
  \subfloat[]{
    \includegraphics[width=0.47\linewidth,pagebox=cropbox,clip]{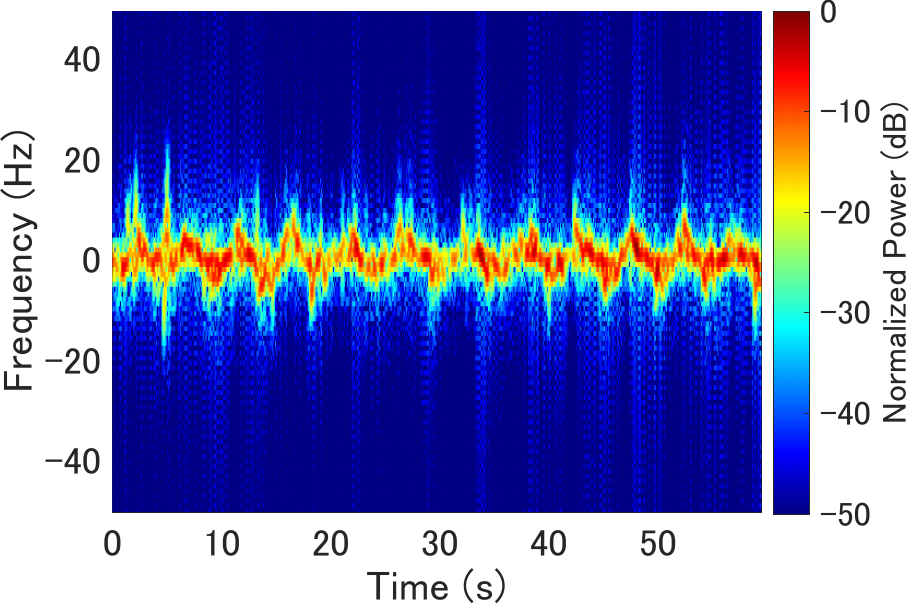}
  }
  \caption{Spectrograms in simulation and measurements; (a) simulation and (b) measurement.}
  \label{fig:sub1DopplerSTFT}
\end{figure}

Finally, the correlation coefficient of the spectrograms $\rho_{\mathrm{S}}$ was 0.699, indicating stable accuracy under all conditions. Figure \ref{fig:sub1DopplerSTFT} compares the spectrograms for subject 1 under C1. The coefficient of correlation between the two spectrograms was 0.708. The comparison reveals that the measured spectrogram had greater temporal variations in power and interference. This is because the simulation assumes a constant reflection power for each scattering center.

In summary, the accuracy evaluations demonstrate that the proposed method successfully reproduces the positions and velocities of scattering centers through simulation.

\begin{table}[tb]
  \centering
  \caption{EVALUATION OF THE PROPOSED SIMULATION METHOD}
  \resizebox{0.95\linewidth}{!}{
    \setlength{\tabcolsep}{0.3em}
    \begin{tabular}{@{}c|cccccc@{}}
            \toprule
      \multirow{2}{*}{Condition} & Radar Image & \multicolumn{2}{c}{Phase} & \multicolumn{2}{c}{Phase HPF} & Spectrogram \\
                           & Corr.       & Corr.     & RMSE (ms)     & Corr.       & RMSE (ms)       & Corr.   \\
                                     \midrule
1.0 m                      & 0.957       & 0.750     & 5.54          & 0.803       & 1.17            & 0.709   \\
2.0 m                      & 0.932       & 0.630     & 6.07          & 0.826       & 1.20            & 0.711   \\
3.0 m                      & 0.969       & 0.208     & 8.92          & 0.667       & 1.56            & 0.702   \\
      \midrule
\ang{45}                         & 0.930       & 0.528     & 3.22          & 0.830       & 0.75            & 0.711   \\
\ang{90}                        & 0.975       & 0.635     & 1.38          & 0.701       & 0.57            & 0.678   \\
\ang{135}                        & 0.922       & 0.599     & 3.52          & 0.532       & 0.55            & 0.691   \\
\ang{180}                        & 0.954       & 0.688     & 2.85          & 0.518       & 0.57            & 0.685   \\
\midrule
2D                         & 0.863       & 0.461     & 4.67          & 0.604       & 1.39            & 0.706   \\
      \midrule
All                        & 0.938       & 0.562     & 4.52          & 0.685       & 0.97            & 0.699  \\
     \bottomrule

    \end{tabular}
  }
  \label{tbl:evalDistance}
\end{table}

\subsection{Comparison with Measurements Made Using a Temperature Sensor}
To verify whether the displacements $d(t)$ obtained from the simulated and measured radar signals originate from respiratory physiological activity, we compared them with measurements made using a reference contact-type temperature sensor. Specifically, we evaluated the similarity between the temperature sensor signal $h(t)$ and the displacements estimated from the simulation $\hat{d}(t)$ and the measurement $d(t)$, using two metrics: the cross-correlation coefficient and the dominant frequency component associated with respiration. The maximum cross-correlation value $\rho_{\mathrm{T}}$ was calculated according to
\begin{equation}
    \rho_{\mathrm{T}} = \max_{\varDelta t} \frac{1}{\Gamma_{\mathrm{h}} \Gamma}_{\mathrm{d}} \int  d_{\mathrm{HF}}(t)h_{\mathrm{HF}}(t+\varDelta t) \, \dd t,
\end{equation}
where $h_{\mathrm{HF}}(t)$ is the temperature sensor signal obtained after applying the high-pass filter described in Section \ref{sec:EVALUATION}.
The respiration rate $f^{\mathrm{RR}}$ is obtained for each signal, and the RMS error across all participants $\epsilon_{\mathrm{RR}}$ and the relative error $\tilde{\epsilon}_{\mathrm{RR}}$ are given by 
\begin{align}
   & f^{\mathrm{RR}} = \underset{f\in \mathcal{E}_{\mathrm{RR}}}{\textrm{argmax}} \; | \mathcal{F}\left[h(t)\right]|^2,\\
   & \epsilon_{\mathrm{RR}} = \sqrt{ \frac{1}{N_{\mathrm{J}}} \sum_{j}^{N_{\mathrm{J}}} |f^{\mathrm{RR}}_j - \hat{f}^{\mathrm{RR}}_j}|^2,                            \\
   & \tilde{\epsilon}_{\mathrm{RR}} =\frac{1}{N_{\mathrm{J}}} \sum_{j}^{N_{\mathrm{J}}} \frac{|f^{\mathrm{RR}}_j- \hat{f}^{\mathrm{RR}}_j|}{f^{\mathrm{RR}}_j},
\end{align}
where $\mathcal{E}_{\mathrm{RR}}$ represents the frequency range for respiration, which was set to $\mathcal{E}_{\mathrm{RR}} = [0.15, 0.4]$ Hz in this study.
$\mathcal{F}$ represents the Fourier transform operator, $j$ is an index indicating the $j$th participant's data, $\hat{f}^{\mathrm{RR}}$ is the respiratory rate estimated from the displacement, and $N_{\mathrm{J}}$ is the total number of data.

\begin{figure}[tb]
  \centering
  \includegraphics[width = 0.7\linewidth,pagebox=cropbox,clip]{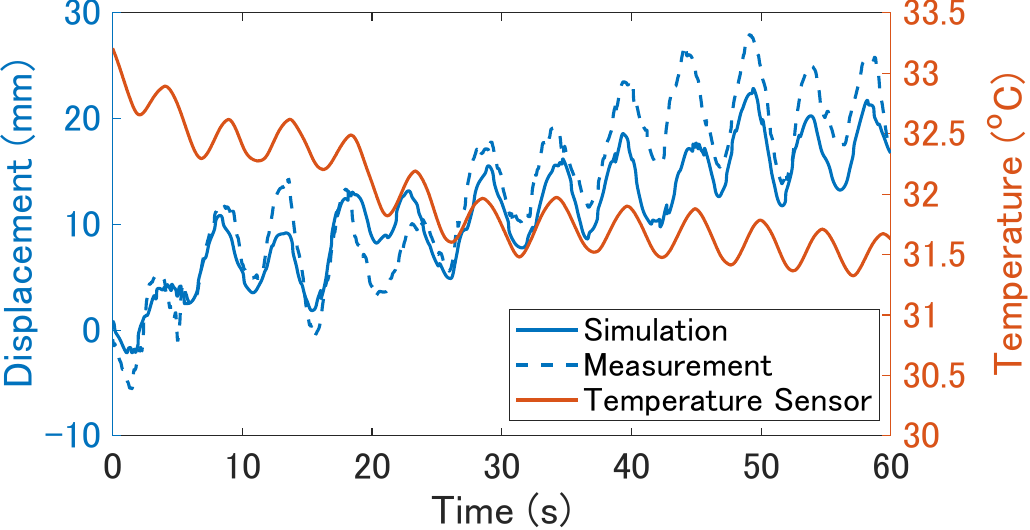}
  \caption{Comparison of the displacements of the radar system and temperature sensor.}
  \label{fig:sub1Temperature}
\end{figure}

Figure \ref{fig:sub1Temperature} compares the temperature sensor signal $h(t)$ and displacement $d(t)$ for participant 1 under C1. It is confirmed that both the measured and simulated displacements agree well with the temperature sensor data. The average results for all participants are summarized in Table \ref{tbl:evalWithTemperature}. Under frontal conditions (C1, C2, C3, C4), both simulation and measurement yielded high accuracy. In particular, the RMS error of the respiratory rate was smaller than $0.005$, which is sufficiently small considering that the frequency resolution for a 60-s measurement is 0.017 Hz. This indicates that both the simulated and measured signals contain respiratory components with sufficient accuracy. By contrast, under back-facing conditions (C6 and C7), the accuracy of the simulation was worse than that of the measurement, indicating that the simulation may be less accurate under back-facing conditions. This can be attributed to the more complex geometry of the human back, which includes features such as the spine and shoulder blades, making electromagnetic scattering simulation more challenging than for the front-facing condition.

\begin{table}[tb]
  \caption{COMPARISON WITH A TEMPERATURE SENSOR}
  \resizebox{1.00\linewidth}{!}{
  \begin{threeparttable}
    \begin{tabular}{c|cccccc}
      \toprule
      \multirow{2}{*}{Condition} & \multicolumn{2}{c}{Cross Corr.} & \multicolumn{2}{c}{RMSE of Freq.   (Hz)} & \multicolumn{2}{c}{Ratio Error of   Freq. \%}                                \\
                                 & Sim.                     & Meas.                                   & Sim.                                    & Meas. & Sim. & Meas. \\
      \midrule
      1.0 m                      & 0.668                          & 0.707                                    & 0.002                                         & 0.003  & 0.74       & 1.15   \\
      2.0 m                      & 0.672                          & 0.706                                    & 0.005                                         & 0.005  & 1.60       & 1.60   \\
      3.0 m                      & 0.606                          & 0.645                                    & 0.005                                         & 0.005  & 1.36       & 1.48   \\ \midrule
      \ang{45}                   & 0.620                          & 0.637                                    & 0.002                                         & 0.003  & 0.72       & 0.90   \\
      \ang{90}                   & 0.519                          & 0.562                                    & 0.029                                         & 0.013  & 5.67       & 2.43   \\
      \ang{135}                  & 0.314                          & 0.448                                    & 0.065                                         & 0.013  & 18.79      & 3.43   \\
      \ang{180}                  & 0.299                          & 0.316                                    & 0.067                                         & 0.013  & 18.13      & 3.73   \\ \midrule
      2D                         & 0.517                          & 0.602                                    & 0.016                                         & 0.005  & 4.52       & 1.71   \\ \midrule
      All                        & 0.527                          & 0.578                                    & 0.035                                         & 0.008  & 6.44       & 2.05   \\
      \bottomrule
    \end{tabular}
          \begin{tablenotes}
      \item[*]Sim.: simulation, Meas.: measurement.
    \end{tablenotes}
  \end{threeparttable}
  }
  \label{tbl:evalWithTemperature}
\end{table}

\subsection{Comparison with Model-Based Displacements}
In this section, we conduct simulations based on model-based displacements as a conventional method and compare the accuracy with that of the proposed method. The conventional method presented in \cite{10.1109/RadarConf2458775.2024.10548317} involves applying model-based displacements to specific body parts of a human phantom. This study adopts a similar approach, manually extracting scattering centers from the torso region of the human body geometry obtained via a depth camera, and applying sinusoidal displacements for simulation. As the geometry is based on measurements from a depth camera rather than a human phantom, the simulation is expected to achieve higher accuracy than the method in \cite{10.1109/RadarConf2458775.2024.10548317}. The distance between each scattering center and the antenna is given as 
\begin{align}
   & R_{m,n}(t) = A_{\mathrm{B}}\sin\left(2\pi f^{\mathrm{RR}} t + \phi_{\mathrm{B}}\right) + \| \bm{p}_m - \bar{\bm{q}}^{\mathrm{B}}_n \|,
  \label{eq:model-baedDisplacement}
\end{align}
where $\bar{\bm{q}}^{\mathrm{B}}_n \in \bar{\mathcal{Q}}_{\mathrm{B}}$ are the scattering centers in the torso, and $A_{\mathrm{B}}$, $f^{\mathrm{RR}}$, and $\phi_{\mathrm{B}}$ represent the amplitude, frequency, and phase of the model-based displacement, respectively. $A_{\mathrm{B}}$ was set at 5 mm and $f^{\mathrm{RR}}$ was determined based on the respiratory rate obtained from the temperature sensor. The value of $\phi_{\mathrm{B}}$ was selected by identifying the phase that maximized the correlation with the displacement $d_{\mathrm{HF}}(t)$ obtained from the measurements.

\begin{figure}[tb]
  \centering
  \begin{minipage}[c]{0.4\linewidth}
    \centering
    \subfloat[]{
      \includegraphics[width=\linewidth,pagebox=cropbox,clip]{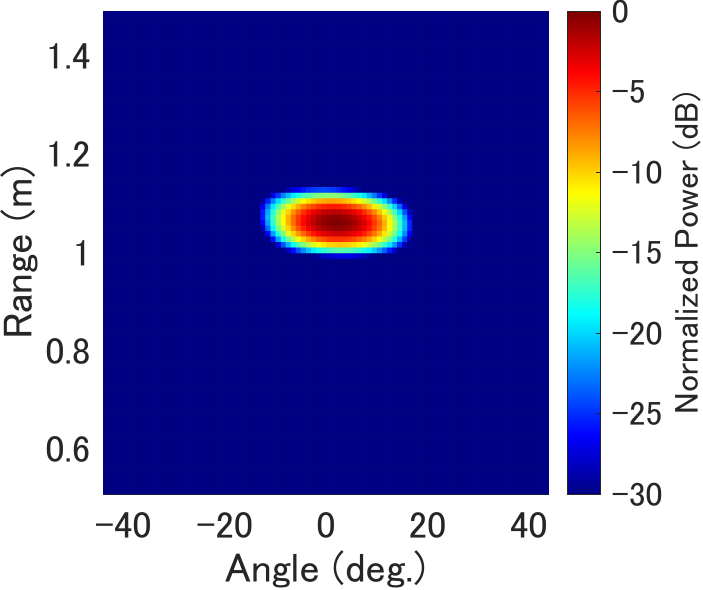}
     \label{fig:C4_ConvMethod_RadarImgage}
      }
  \end{minipage}
  \hfill
  \begin{minipage}[c]{0.55\linewidth}
    \centering
    \subfloat[]{
      \includegraphics[width=\linewidth,pagebox=cropbox,clip]{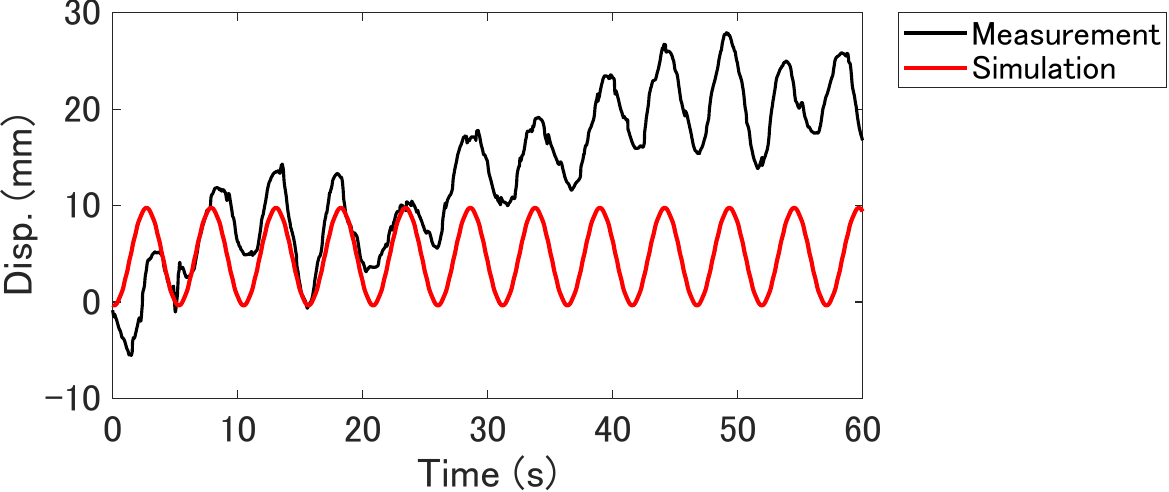}
       \label{fig:C4_ConvMethod_PhaseDiplacement}
    }
  \end{minipage}
  \\
  \subfloat[]{
    \includegraphics[width=0.47\linewidth,pagebox=cropbox,clip]{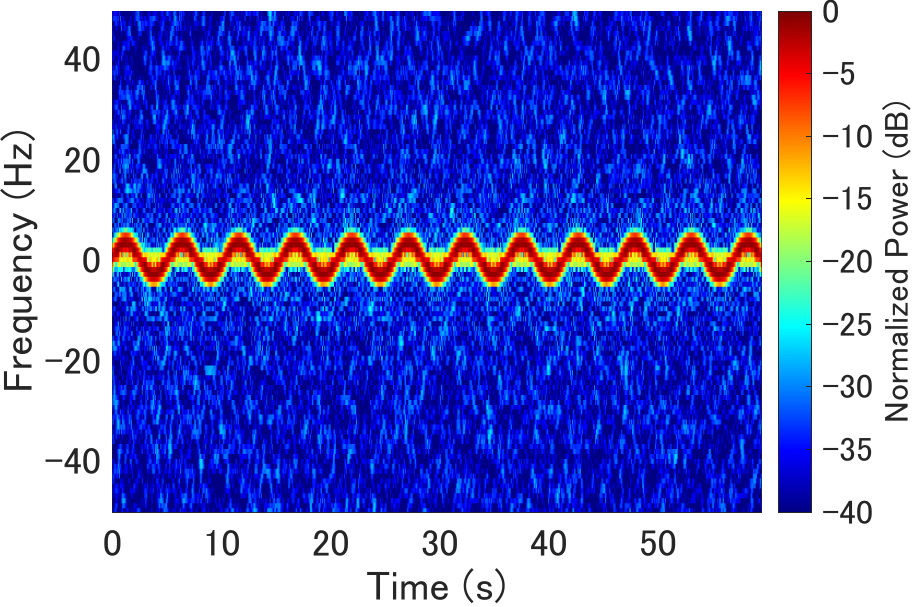}
      \label{fig:C4_ConvMethod_Doppler}
  }

  \caption{(a) Radar image, (b) displacement, and (c) spectrogram for the conventional method.}
  \label{fig:C4_ConvMethod}
\end{figure}

Figure~\ref{fig:C4_ConvMethod} shows the results of the conventional method for the radar image, displacement, and spectrogram under C1 for subject 1. The coefficients of correlation with the measurements were 0.793 for the radar image, 0.359 for the displacement, and 0.681 for the spectrogram. Conversely, the proposed method achieved correlation coefficients of 0.965, 0.936, and 0.708, indicating a substantial improvement in accuracy.
A comparison of the radar images produced by the conventional and proposed methods reveals that the conventional method (Fig. ~\ref{fig:C4_ConvMethod_RadarImgage}) applies displacement only to the torso, resulting in a single echo. By contrast, the proposed method (Fig. ~\ref{fig:C4_RadarImageSub1C1Sim}) applies displacement to scattering centers derived from electromagnetic scattering, enabling more accurate simulation of echoes from the human body.
A comparison of the results of phase displacement with those of the conventional model-based approach (Figs. ~\ref{fig:conv_mukherjee} and ~\ref{fig:C4_ConvMethod_PhaseDiplacement}) shows that the phase displacement generated using the proposed method (Fig. ~\ref{fig:sub1PhaseDisp}) closely aligns with the measured displacement.
Notably, the proposed method does not use any information from contact sensors and radar measurements, unlike the conventional method. Despite this, the proposed method successfully generates the IF signals that correlate with the measurements.

Table \ref{tbl:evalWithConventional} summarizes the average accuracy of the conventional method for all participants under C1. Compared with the results obtained using the proposed method in Table \ref{tbl:evalDistance}, improvements are observed on all evaluation metrics. Specifically, the correlation coefficients of the radar image, displacement, and spectrogram improved by 7.5\%, 58.2\%, and 3.2\%, respectively, demonstrating that the proposed method is particularly effective in generating simulated displacement signals that closely resemble measurements.

 \begin{table}[tb]
  \centering
  \caption{COMPARISON WITH THE CONVENTIONAL METHOD}
  \resizebox{0.95\linewidth}{!}{
    \setlength{\tabcolsep}{0.3em}
    \begin{tabular}{@{}c|cccccc@{}}
            \toprule
      \multirow{2}{*}{} & Radar Image & \multicolumn{2}{c}{Phase} & \multicolumn{2}{c}{Phase HPF} & Spectrogram \\
                           & Corr.       & Corr.     & RMSE (ms)     & Corr.       & RMSE (ms)       & Corr.   \\
                                     \midrule
Conv.                 & 0.890       & 0.474     & 5.65          & 0.709       & 2.60           & 0.687   \\
     \bottomrule
    \end{tabular}
  }
  \label{tbl:evalWithConventional}
\end{table}

\section{Conclusion}
In this study, we proposed a method for simulating the received signal of an FMCW radar system for respiratory measurement using human body geometry obtained with a depth camera. The proposed approach enables signal simulation directly from depth camera measurements, without relying on physical phantoms or model-based motion, by leveraging a scattering center model based on high-frequency electromagnetic scattering theory. To evaluate the performance of the method, experiments were conducted with six participants under varying conditions, including varying target distances, seating directions, and radar types. The average correlation coefficient was 0.938 for radar images and 0.699 for spectrograms. A comparative analysis with a conventional model-based displacement method demonstrated improvements of 7.5\%, 58.2\%, and 3.2\% in radar image similarity, target displacement similarity, and spectrogram similarity, respectively. This study contributes to the generation of simulated radar signals for respiratory monitoring and provides insight into the factors affecting accuracy in non-contact sensing under diverse experimental conditions. Future work will include modeling the time-varying echo power and extending the framework to simultaneously simulate both respiration and heartbeat signals.

\section*{Ethics Declarations}
The experimental protocol involving human participants was approved by the Ethics Committee of the Graduate School of Engineering, Kyoto University (permit no. 202223). Informed consent was obtained from all human participants in the study.

\section*{Acknowledgment}
The authors report no conflict of interest. This work was supported in part by JST SPRING Grant JPMJSP2110; the SECOM Science and Technology Foundation; in part by the Japan Science and Technology Agency under Grant JPMJMI22J2 and Grant JPMJMS2296; in part by the Japan Society for the Promotion of Science KAKENHI under Grant 21H03427, Grant 23H01420, and Grant 23K26115; and in part by the New Energy and Industrial Technology Development Organization. The authors thank Glenn Pennycook, MSc, from Edanz (https://jp.edanz.com/ac) for editing a draft of this manuscript.

%%% Bibliography

%%%% Biography
\begin{IEEEbiography}[{\includegraphics[width=1in,height=1.25in,clip,keepaspectratio]{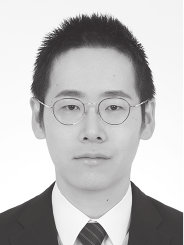}}]
  {Kimitaka Sumi} received B.E. and M.E. degrees from Nagoya University, Aichi, Japan, in 2020 and 2022, respectively. He is currently studying for a Ph.D. degree at the Graduate School of Engineering, Kyoto University. His research interests include signal processing and remote sensing.
\end{IEEEbiography}
\begin{IEEEbiography}[{\includegraphics[width=1in,height=1.25in,clip,keepaspectratio]{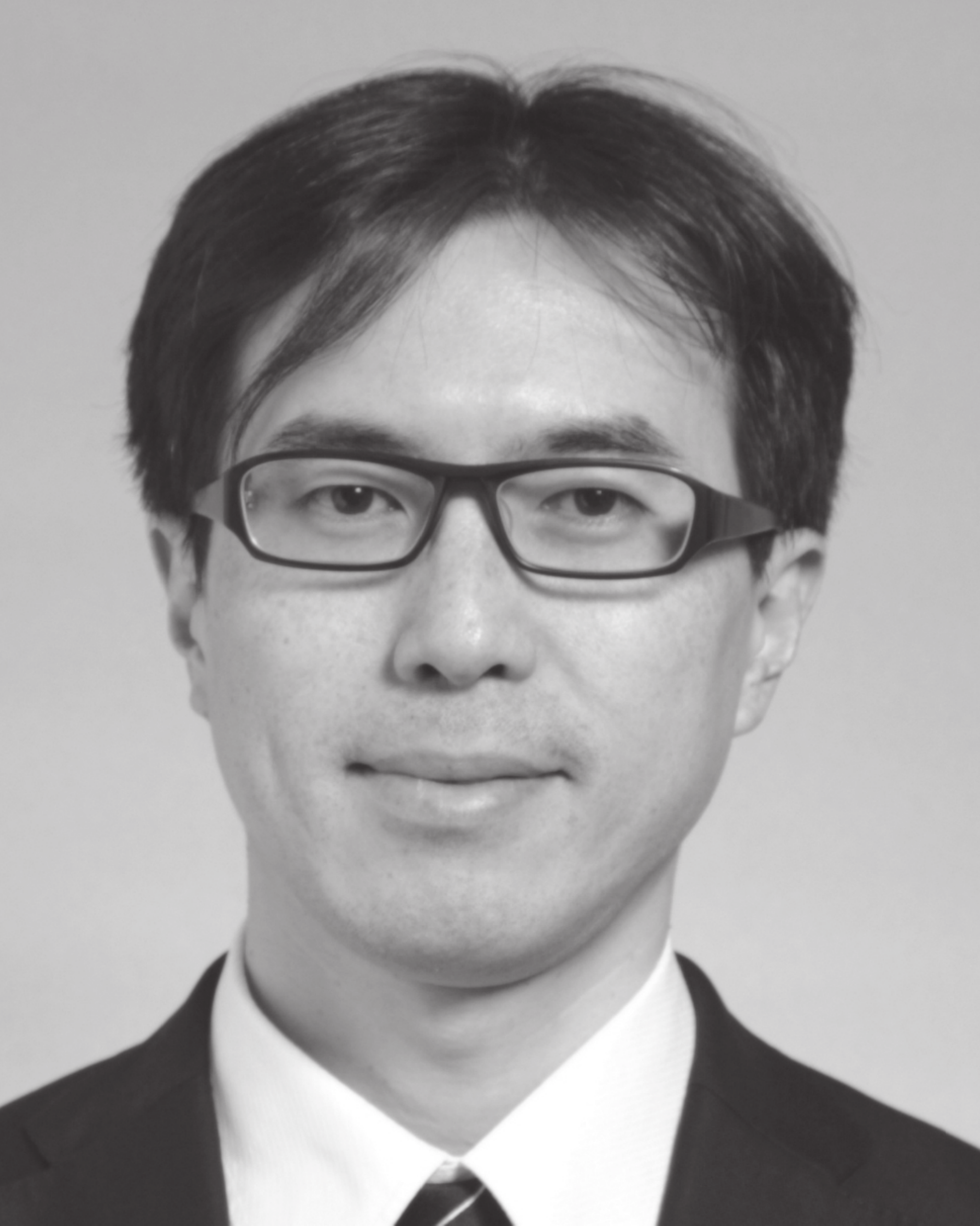}}]
  {Takuya Sakamoto} (Senior Member, IEEE)
  received a B.E. degree in electrical and electronic engineering from Kyoto University, Kyoto, Japan, in 2000 and M.I. and Ph.D. degrees in communications and computer engineering from the Graduate School of Informatics, Kyoto University, in 2002 and 2005, respectively. From 2006 through 2015, he was an Assistant Professor at the Graduate School of Informatics, Kyoto University. From 2011 through 2013, he was also a Visiting Researcher at Delft University of Technology, Delft, the Netherlands. From 2015 until 2019, he was an Associate Professor at the Graduate School of Engineering, University of Hyogo, Himeji, Japan. In 2017, he was also a Visiting Scholar at the University of Hawaii at Manoa, Honolulu, HI, USA. From 2019 until 2022, he was an Associate Professor at the Graduate School of Engineering, Kyoto University. From 2018 through 2022, he was a PRESTO researcher of the Japan Science and Technology Agency, Japan. Since 2022, he has been a Professor at the Graduate School of Engineering, Kyoto University. His current research interests lie in wireless human sensing, radar signal processing, and radar measurement of physiological signals.

  Prof. Sakamoto was a recipient of the Best Paper Award from the International Symposium on Antennas and Propagation (ISAP) in 2004, the Young Researcher's Award from the Institute of Electronics, Information and Communication Engineers of Japan (IEICE) in 2007, the Best Presentation Award from the Institute of Electrical Engineers of Japan in 2007, the Best Paper Award from the ISAP in 2012, the Achievement Award from the IEICE Communications Society in 2015, 2018, and 2023, the Achievement Award from the IEICE Electronics Society in 2019, the Masao Horiba Award in 2016, the Best Presentation Award from the IEICE Technical Committee on Electronics Simulation Technology in 2022, the Telecom System Technology Award from the Telecommunications Advancement Foundation in 2022, and the Best Paper Award from the IEICE Communication Society in 2007 and 2023.
\end{IEEEbiography}

\end{document}